\documentclass[floats,floatfix,amssymb,prd,twocolumn,superscriptaddress,nofootinbib,preprintnumbers]{revtex4-1}

\usepackage{amssymb,amsmath,verbatim,mathtools,needspace,enumitem,etoolbox,graphicx,physics,microtype,afterpage,bm}
\usepackage[dvipsnames, usenames]{xcolor}
\definecolor{linkcolor}{rgb}{0.0,0.3,0.5}
\usepackage[unicode, colorlinks=true, linkcolor=linkcolor, citecolor=linkcolor, filecolor=linkcolor,urlcolor=linkcolor, pdfusetitle]{hyperref}
\usepackage[all]{hypcap}
\usepackage[T1]{fontenc}
\usepackage[utf8]{inputenc}
\usepackage{tabularx}
\usepackage{float}
\usepackage{psfrag}

\usepackage{slashed}
\usepackage{lmodern}
\allowdisplaybreaks
\usepackage{tikz}
\usepackage{color}
\usepackage{framed}
\usepackage{hyperref}
\hypersetup{colorlinks, citecolor=bluscuro, linkcolor=black, urlcolor=bluscuro}
\definecolor{rossos}{cmyk}{0,1,1,0.55}
\definecolor{bluscuro}{rgb}{0.15, 0.2, .85}
\definecolor{bluchiaro}{cmyk}{1,.3,0.,0.1}
\definecolor{ForestGreen}{rgb}{0.13, 0.55, 0.13}

\def\lsim{\mathrel{\rlap{\lower4pt\hbox{\hskip0.5pt$\sim$}}
    \raise1pt\hbox{$<$}}}         
\def\gsim{\mathrel{\rlap{\lower4pt\hbox{\hskip0.5pt$\sim$}}
    \raise1pt\hbox{$>$}}}         

\begin{document}

\title{Message in a bottle:\\
energy extraction from bouncing geometries}
\author{Vitor Cardoso}
\affiliation{Niels Bohr International Academy, Niels Bohr Institute, Blegdamsvej 17, 2100 Copenhagen, Denmark}
\affiliation{CENTRA, Departamento de F\'{\i}sica, Instituto Superior T\'ecnico -- IST, Universidade de Lisboa -- UL,
Avenida Rovisco Pais 1, 1049 Lisboa, Portugal}

\author{Jo\~ao L. Costa}
\affiliation{Departamento de Matem\'{a}tica, Instituto Universit\'ario de Lisboa (ISCTE-IUL), Av. das For\c{c}as Armadas, 1649-026 Lisboa, Portugal,} 
\affiliation{CAMGSD, Instituto Superior T\'ecnico -- IST, Universidade de Lisboa -- UL, Avenida Rovisco Pais 1, 1049-001, Lisboa, Portugal}

\author{Jos\'e Nat\'ario}
\affiliation{Departamento de Matem\'{a}tica, Instituto Superior T\'ecnico -- IST, Universidade de Lisboa -- UL, Avenida Rovisco Pais 1, 1049-001, Lisboa, Portugal}
\affiliation{CAMGSD, Instituto Superior T\'ecnico -- IST, Universidade de Lisboa -- UL, Avenida Rovisco Pais 1, 1049-001, Lisboa, Portugal}

\author{Zhen Zhong}
\affiliation{CENTRA, Departamento de F\'{\i}sica, Instituto Superior T\'ecnico -- IST, Universidade de Lisboa -- UL,
Avenida Rovisco Pais 1, 1049 Lisboa, Portugal}

\begin{abstract}
Quantum gravity phenomenology suggests the interesting possibility that black holes are not eternal. Collapse could be halted by some unknown mechanism, or Hawking radiation might leave behind a regular spacetime. Here we investigate a simple bouncing geometry, with (outer and inner) apparent horizons but no event horizon. We show that the inner horizon blueshifts radiation, which can lead to a gigantic amplification of energy observable from far away regions. Thus, if such phenomena exists in our universe, they can power high-energy bursts at late stages in their lives, when the horizons disappear and spacetime bounces back to a flat geometry.
\end{abstract}

\maketitle

\noindent {\bf \em Introduction.} 
Hawking radiation and the consequent black hole evaporation are among the most remarkable possibilities raised by theoretical physics in the last half century. 
These are prime examples of quantum gravitational phenomena, but, due to the lack of a fully established theory of quantum gravity, their understanding stems mostly from semiclassical arguments (as the ones that led to their discovery~\cite{Hawking:1974rv}), where gravity is treated classically and matter quantum mechanically. Similar phenomena arise also from somewhat informal pictures developed in the context of candidates to full quantum gravity theories, such as Loop Quantum Gravity~\cite{Ashtekar:2023cod}, and from phenomenological approaches leading to classical metrics that explicitly capture some of the features suggested by the proposals above~\cite{Hayward:2005gi}.

\begin{figure}[h!]
\centering
\includegraphics[width=0.6\linewidth]{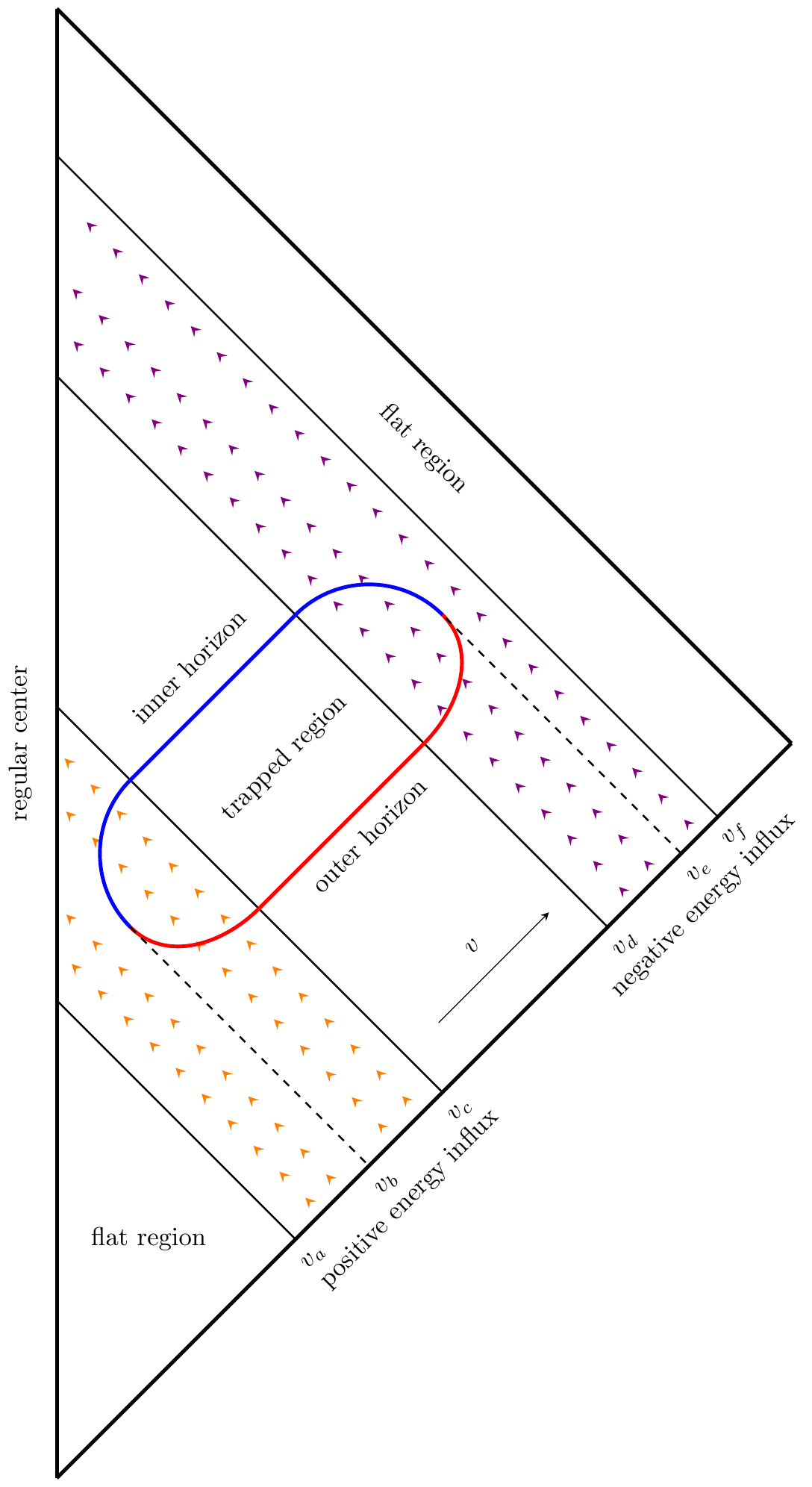}
    \caption{Penrose diagram of a (everywhere regular) bouncing geometry. Following the advanced time coordinate $v$, we start ($v<v_a$) with a flat region; then, an influx ($v_a<v<v_c$) of positive energy leads to the formation ($v=v_b$) of two horizons bounding a region containing trapped surfaces; afterwards, an influx of negative energy ($v_d<v<v_f$) leads to the evaporation of the horizons ($v=v_e$), and to a flat spacetime in the far future region ($v>v_f$).}
    \label{fig:penrose}
\end{figure}
Even under this theoretical uncertainty, and without observational evidence, it is widely accepted that some form of radiation and evaporation play an essential role in black hole dynamics. It is then natural to try to understand what distinguishes such quantum gravitational objects from their classical counterparts, and, more excitingly, if any of these differences might show up under observational scrutiny.

Here, we consider a simple family of bouncing geometry models, prescribed by classical metrics which are simplified versions of a proposal by Hayward~\cite{Hayward:2005gi}, describing fully regular spacetimes containing a ``black hole'' that forms dynamically and later on evaporates. The Penrose diagram of such a dynamical spacetime is depicted in Fig.~\ref{fig:penrose}. Strictly speaking, these spacetimes don't contain a black hole region, since every event can be seen from future null infinity. Nonetheless, they have a trapped region bounded by two apparent horizons, an inner horizon and an outer horizon, which form dynamically and eventually evaporate, leaving behind a flat spacetime in the far future. 

Our goal is to understand how these spacetimes respond to perturbations, by re-evaluating the  geometric optics approximation and, more importantly, by studying the evolution of (massless) scalar fields.  
We show that the evaporation and consequent disappearance of the horizons leaves behind a clear signature: a burst of extremely high energy generated at the inner horizon. This burst may lead to detectable observational signals, either in gravitational waves or in the electromagnetic spectrum, that could allow us to gain further insight into the quantum nature of astrophysical horizons. 

\noindent {\bf \em A dynamical ``bouncing'' spacetime.} 
We consider a spherically symmetric metric of the form 
\begin{equation} \label{metric}
d s^2=- F(v, r) d v^2 + 2 dv dr + r^2 d \Omega^2 \, ,
\end{equation}
where $d\Omega^2$ is the round metric on the unit sphere and
\begin{equation}
F(v, r)=1-\frac{2 m(v) r^2}{r^3+2 l^2 m(v)}\,.
\end{equation}
Here, the constant $l$ is assumed to be positive, and the mass function $m(v)$ is assumed to be non-negative. There are two positive real solutions $r_{\pm}(v)$ of the equation $F(v,r)=0$ when $m(v) > m_* = \frac{3 \sqrt{3}}{4}l$, corresponding to an inner and an outer apparent horizons.

We are interested in describing a spacetime which is flat at early and late times, but has transient horizons.
We follow Ref.~\cite{Hayward:2005gi} in choosing $m(v)$ to be a smooth function that (nearly) vanishes in the intervals $(-\infty, v_a)$ and $(v_f, \infty)$, is (nearly) constant equal to $m_0>m_*$ in the interval $(v_c, v_d)$, is increasing in the interval $(v_a, v_c)$, and is decreasing in the interval $v\in(v_d, v_f)$, where $v_a<v_c<v_d<v_f$ are adjustable parameters. Under these assumptions, the horizons are formed at some advanced time $v_b \in (v_a,v_c)$ and disappear at $v_e \in (v_d,v_f)$, where $v_b$ and $v_e$ are determined by $m(v_b)=m(v_e)=m_*$. 
Concretely, we choose
\begin{equation}
\label{eq:mass_function}
m(v) = \frac{m_0}{2} \left[ \tanh(s\frac{v-v_a}{v_c-v_a}) - \tanh(s\frac{v-v_d}{v_f-v_d}) \right] \, ,
\end{equation}
which closely resembles Hayward's profile.
In particular, for larger values of $s$, the slope in the transition intervals $(v_a,v_c)$ and $(v_d,v_f)$ becomes steeper. For concreteness, here we take $s = 10$.

\noindent {\bf \em Blueshift of null geodesics.} 
%
\begin{figure}
    \centering
    \includegraphics[width=\linewidth]{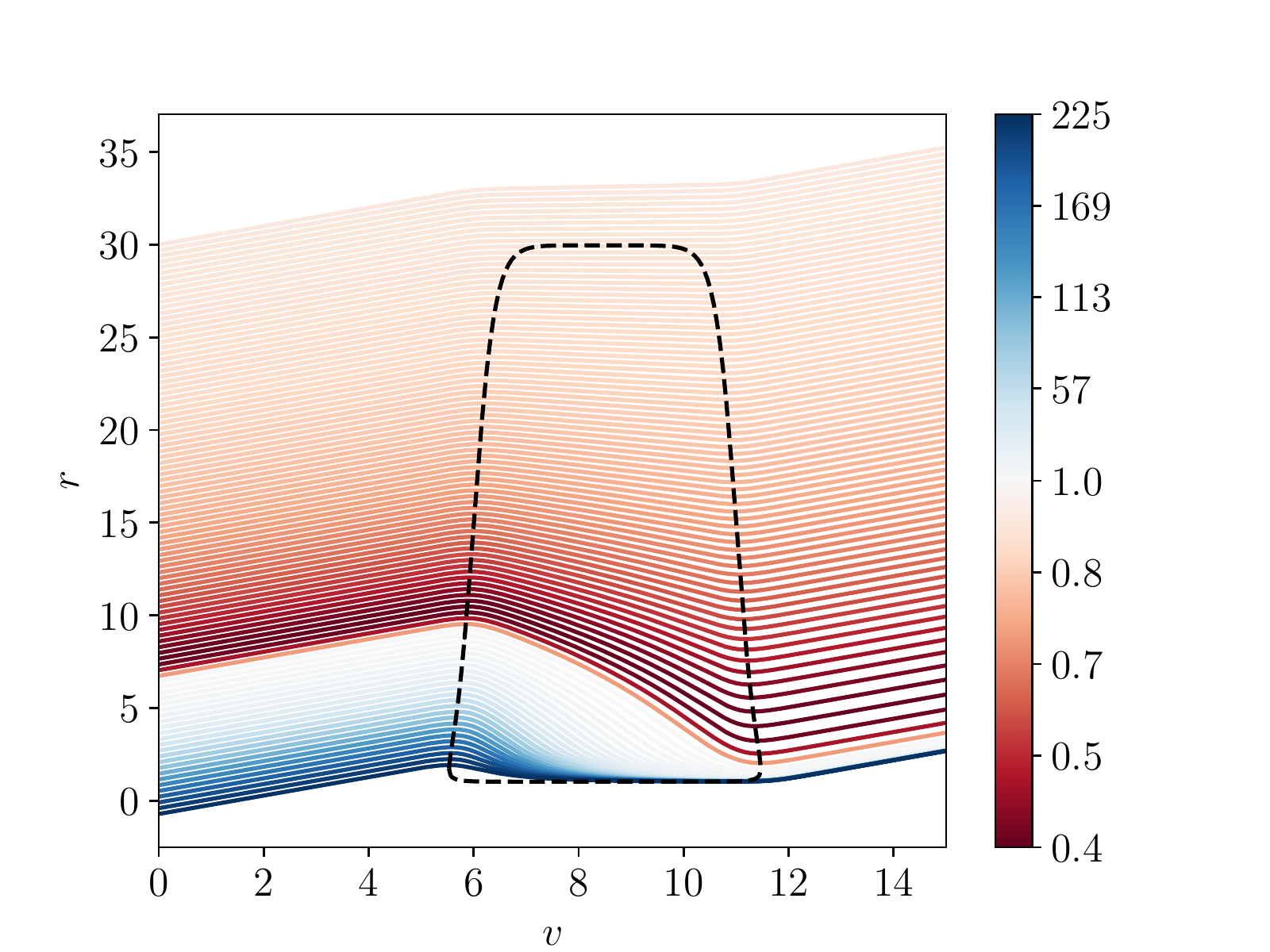}
    \caption{Null geodesics for spacetime \texttt{IA}. The black dashed line shows the location of the horizons. Notice how null geodesics pile up near the inner horizon, which is ultimately the reason for the high-energy burst of radiation that we see once the spacetime bounces and horizons disappear, cf. Fig.~\ref{fig:scalar}. The line color indicates the amplification factor, with darker blue lines representing a stronger blueshift and darker red lines indicating a stronger redshift. In regions located far away from the inner horizon, there is only a minimal amount of redshift.}
    \label{fig:null_geodesic}
\end{figure}
Our main result is a large blueshift of the radiation trapped inside the horizons. 
This radiation may have been absorbed during horizon formation, or accreted at later stages. 
Although our focus is on massless (scalar) fields, we can see the blueshift already at the level of null geodesics, a good description of high-frequency radiation.
This geometric optics approximation was previously studied in Ref.~\cite{Frolov:2017rjz}, although for a different mass function. We quickly revisit this problem for the mass function~\eqref{eq:mass_function}, in order to compare with our main result concerning the amplification of scalar waves.
Consider a radially outgoing light ray. The energy of the corresponding photon with respect to stationary observers in the flat region, as well as observers at the center $r=0$ (for whom $F(v,r)=1$), is simply
\begin{equation}
E = - \frac{\partial L}{\partial \dot{v}} = F(v,r)\dot{v}-\dot{r} \,,
\end{equation}
where
\begin{equation}
L = \frac12 \left[ - F(v, r) \dot{v}^2 + 2 \dot{v} \dot{r}\right]
\end{equation}
is the Lagrangian for radial geodesics. This is also a conserved quantity in the static Schwarzschild-like region where $m$ is constant, and so we identify it with the photon's energy. 
For outgoing null geodesics, we have
\begin{equation} \label{null_geo}
F(v, r) \dot{v} = 2\dot{r} \quad \Rightarrow \quad  
\frac{dr}{dv} = \frac12 F(v, r) \, ,
\end{equation}
and also
\begin{equation} \label{energy}
E = \dot{r} = \frac12 F(v, r) \dot{v} \, . 
\end{equation}
Using the Euler-Lagrange equation in $r$, we find that
\begin{equation}
\ddot{v} = - \frac12 \frac{\partial F}{\partial r} \dot{v}^2 \,,
\end{equation}
and so, assuming $v=v_0$ and $\dot{v}=\dot{v}_0$ for $r=0$, we have
\begin{equation}
\dot{v} = \dot{v}_0 \exp\left( - \frac12 \int_{v_0}^v \frac{\partial F}{\partial r} dv \right) \,,
\end{equation}
where $r(v)$ in this integral is computed from Eq.~\eqref{null_geo}. From Eq.~\eqref{energy} we then obtain the ratio between the energies of the photon at $r=0$ and $r=\infty$:
\begin{equation}
\mathcal{A}\equiv \frac{E_\infty}{E_0} = \exp \left( - \frac{1}{2} \int_{v_0}^\infty \frac{\partial F}{\partial r} dv \right) \, .
\label{eq:amplification_geodesics}
 \end{equation}

We evaluated the integral in \eqref{eq:amplification_geodesics} to compute the largest and smallest amplification factor $\mathcal{A}$ for different spacetimes, shown in Table~\ref{table:results} and Fig.~\ref{fig:null_geodesic}.
Notice that the amplification can become arbitrarily large for sufficiently long-lived horizons (when $v_d-v_c$ is very large compared to the other scales in the problem), and stems from geodesics close to the inner horizon, which is in fact an attractor for radial null geodesics, as is easily seen from Eq.~\eqref{null_geo}.

 In the limit $v_a\sim v_c$, $v_d\sim v_f$, and for null geodesics that stay close to the inner horizon, we have the approximate formula
 \begin{equation}
\mathcal{A} \sim \exp \left( \kappa_- (v_d-v_c) \right) \, ,\label{amplification_geo}
 \end{equation}
where
\begin{equation}
\kappa_- = - \frac{m_0 r_-^4 - 4l^2m_0^2r_-}{\left(r_-^3+2l^2m_0\right)^2}\,,
\end{equation}
is the surface gravity of the inner horizon, with $r_-$ the smallest positive root of
$r_-^3-2 m_0 r_-^2+2 l^2 m_0$.
\begin{table}
\caption{List of simulations for the evolution of the $\ell=0$ mode of a scalar field in spacetime~\eqref{metric}-\eqref{eq:mass_function}. We use $r_G = 8$, $\sigma = 1$ for the initial pulse, with initial energy $E_0 = 0.63$. We also use $s = 10$ for the mass function $m(v)$. We define the amplification $\mathcal{A}\equiv E_\infty/E_0$, the ratio between final and initial energy of the scalar field, cf. Eq.~\eqref{eq:E}. The quantities $\mathcal{A}_g^{\rm max}$, $\mathcal{A}_g^{\rm min}$, are the corresponding amplification factors for null geodesics which maximize or minimize the energy gain, respectively, cf. Eq.~\eqref{eq:amplification_geodesics}. We get similar results for $\ell >0$. \label{table:results}}
\begin{ruledtabular}
\begin{tabular}{cccccccccccccc}
Run         & $m_0$ & $v_a$  & $v_c$ &$v_d$ & $v_f$ & $l$ & $\kappa_-$ & $E_f$ & $\mathcal{A}$ & $\mathcal{A}_g^{\rm max}$ & $\mathcal{A}_g^{\rm min}$   \\
\hline
\texttt{IA}   & $15$   & $6$  & $10$   & $11$  & $15$  & $1$ & $0.9329$ & $4.55$ & $7.26$ & $225$ & $0.38$  \\
\texttt{IIA}  & $15.5$ & $6$  & $10$   & $11$  & $15$  & $1$ & $0.9351$ & $5.15$ & $8.22$ & $232$ & $0.38$  \\
\texttt{IIIA} & $16$   & $6$  & $10$   & $11$  & $15$  & $1$ & $0.9371$ & $5.67$ & $9.06$ & $239$ & $0.38$  \\
\texttt{IVA}  & $16.5$ & $6$  & $10$   & $11$  & $15$  & $1$ & $0.9390$ & $6.10$ & $9.73$ & $246$ & $0.37$ \\
\texttt{VA}   & $17$   & $6$  & $10$   & $11$  & $15$  & $1$ & $0.9408$ & $6.40$ & $10.2$ & $253$  & $0.37$  \\
\texttt{ILA}  & $15$   & $6$  & $10$   & $11$  & $15$  & $2$ & $0.4324$ & $0.47$ & $0.747$ & $10.7$ & $0.63$  \\
\texttt{ILB}  & $15$   & $6$  & $10$   & $11$  & $15$  & $4$ & $0.1813$ & $0.63$ & $0.998$ & $2.59$ & $0.82$   \\
\texttt{IMA}  & $15$   & $6$  & $10$   & $12$  & $16$  & $1$ & $0.9329$ & $11.3$ & $18.1$  & $573$  & $0.34$   \\
\texttt{IMB}  & $15$   & $6$  & $10$   & $13$  & $17$  & $1$ & $0.9329$ & $26.5$ & $42.3$  & $1457$ & $0.31$   \\
\texttt{IMC}  & $15$   & $6$  & $10$   & $14$  & $18$  & $1$ & $0.9329$ & $69.0$ & $110$   & $3704$ & $0.28$   \\
\texttt{IMD}  & $15$   & $6$  & $10$   & $15$  & $19$  & $1$ & $0.9329$ & $178$ & $285$    & $9416$ & $0.26$   \\
\texttt{IME}  & $15$   & $6$  & $10$   & $16$  & $20$  & $1$ & $0.9329$ & $459$ & $733$    & $23935$ & $0.24$  \\
\end{tabular}
\end{ruledtabular}
\end{table}
Note that for $l\ll m_0$ we have $r_-\sim l$ and $\kappa_-\sim 1/l$, and so the amplification acquires the simple form $\mathcal{A} \sim \exp \left((v_d-v_c)/l \right) $.

\noindent {\bf \em Dynamics of a massless scalar field.} 
To verify that energy extraction holds also for large wavelength fields, a proper description of radiation is necessary. For simplicity, we follow the dynamics of a massless scalar field on the above background, governed by the wave equation
\begin{equation}
\nabla^a \nabla_a \Phi = 0\,.
\end{equation}
We can expand $\Phi$ as a superposition of spherical modes, given in terms of the spherical harmonics $Y_{\ell m}(\theta, \varphi)$ as solutions of the form
\begin{equation}
\Phi(v, r, \theta, \varphi) = \frac{\phi(v, r)}{r} Y_{\ell m}(\theta, \varphi)\, ,
\end{equation}
and so we focus on these solutions. The wave equation for the modes can be written explicitly as follows:
\begin{equation}
\partial_v\partial_r \phi + \frac{F}{2} \partial_r^2 \phi + \frac{\partial_r F}{2}\partial_r\phi - \left(\frac{\partial_r F}{2r} + \frac{\ell (\ell + 1)}{2r^2}\right)\phi = 0 \,. \label{eq:eom}
\end{equation}
In what follows we will solve this equation numerically as a characteristic initial value problem by giving initial data on the characteristic surface $v=v_a$. Specifically, we will take a Gaussian centered at $r=r_G$ with width $\sigma$:
\begin{align}
\phi\left(r, v_a\right) &=e^{-\left(\frac{r-r_G}{\sigma}\right)^2}\,.\label{eq:initdata}
\end{align}

Even though the spacetime is dynamic, it has static regions where one has well-defined notions of energy. The energy of the field on a characteristic surface $\mathcal{N}$ of constant $v$ contained in the static regions can be computed as
\begin{equation}
E = - \int_{\mathcal{N}} T_{\mu\nu} \left(\frac{\partial}{\partial v}\right)^\mu \left(\frac{\partial}{\partial r}\right)^\nu= - \int_0^{\infty} \!\!\!\! \int_{S^2} T_{vr} r^2 d\Omega dr \,.
\end{equation}
From the usual expression for the massless scalar field energy-momentum tensor, we find
\begin{equation}
T_{vr} = - \frac12 F (\partial_r \Phi)^2 - \frac1{2r^2} |\slashed{\nabla} \Phi|^2\, ,
\end{equation}
where $\slashed\nabla$ is the gradient on the unit $2$-sphere. Since the spherical harmonics $Y_{\ell m}(\theta, \varphi)$ form an orthonormal basis of eigenfunctions of the Laplacian on the unit $2$-sphere with eigenvalues $-\ell(\ell+1)$, we obtain, after integrating by parts,
\begin{equation}
E = \frac12 \int_0^{\infty} \left[ F r^2 \left( \partial_r\left(\frac{\phi}r\right) \right)^2 + \frac{\ell(\ell+1)}{r^2} \phi^2 \right] dr \,.\label{eq:E}
\end{equation}
We define the amplification factor $\cal{A}$ as we did for the null geodesics, $\mathcal{A}\equiv E_\infty/E_0$.

\noindent {\bf \em Numerical scheme.} 
We can proceed to solve Eq.~\eqref{eq:eom} by employing numerical methods. Specifically, we apply a 2$^{\rm nd}$ order finite difference method for spatial discretization and a 4$^{\rm th}$ order Runge-Kutta method for time integration.
We evolve $\phi$ and $\partial_r \phi$ using the boundary conditions
\begin{equation}
\begin{cases}
\phi(v,0) = 0 \qquad\qquad\qquad\quad \text{(regularity at the origin)\;,} \\
\phi(v,\infty) = \partial_r\phi(v,\infty) = 0 \quad \text{(no incoming radiation)\;,}
\end{cases}\nonumber
\end{equation}
together with the implicit boundary condition obtained by integrating Eq.~\eqref{eq:eom}:
\begin{equation} \label{missing_bc}
\partial_r\phi(v,0) = - \int_0^{\infty} \left(\frac{\partial_r F}{r} + \frac{\ell(\ell + 1)}{r^2} \right) \phi \, dr \,.
\end{equation}
More precisely, given $\phi$ and $\partial_r \phi$ on a surface of constant $v$, we recompute $\partial_r\phi(v,0)$ from Eq.~\eqref{missing_bc} (if $\ell > 0$, the convergence of the integral in Eq.~\eqref{missing_bc} implies immediately that $\partial_r \phi(v, 0) = 0$), which we use to evaluate $\partial^2_r\phi$. We then have $\partial_v \partial_r \phi$ from Eq.~\eqref{eq:eom}, and $\partial_v \phi$ from
\begin{equation}
\partial_v \phi(v,r) = -\int_{r}^{\infty} \partial_v \partial_r \phi \, dr \,,
\end{equation}
which we use to evolve $\phi$ and $\partial_r \phi$ by the method of lines.

In practice, we use a finite computational domain rather than an infinite one.
Furthermore, in order to improve the resolution at the inner horizon, we introduce a new radial coordinate $R$ by stretching $r$ as (see~\cite{Pollney:2009yz})
\begin{align}
f(R) &=A\left(R-R_0\right)+B \sqrt{1+\left(R-R_0\right)^2 / \epsilon} \,,\\*
r &= f(R) - f(0) \,.
\end{align}
By using these coordinates, it is possible to smoothly transition between the two resolutions. These resolutions are determined by the values of the parameters $A$ and $B$, and the transition takes place in a region with a width of $\epsilon$ centered around the value of $R_0$. To be specific, we use $R_0 = 1000$, $\epsilon = 10$, and obtain the values of $A$ and $B$ by solving $r'(0) = 0.01$ and $\lim_{R \to \infty} r'(R) = 1$.
This allows us to resolve sharply shaped waveforms caused by blueshift near the inner horizon. Our numerical results show second-order convergence, consistent with the scheme we use.
%

\noindent {\bf \em Blueshift of scalar fields.} 
%
\begin{figure}
    \centering
    \includegraphics[width=\linewidth]{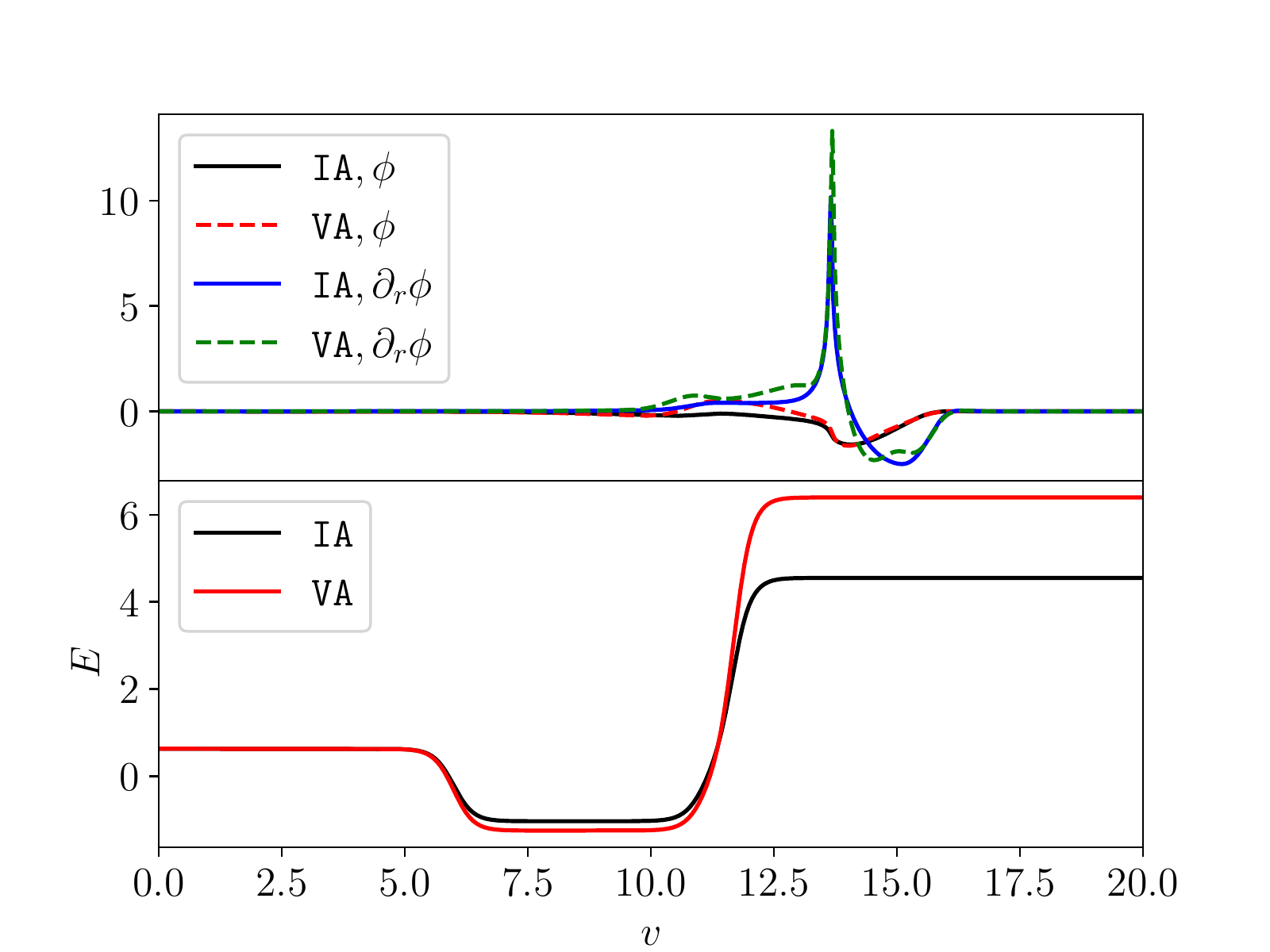}
    \caption{Evolution of Gaussian initial data in bouncing spacetimes. {\bf Top panel:} the scalar field measured at $r \approx 2.0$ for spacetimes \texttt{IA} and \texttt{VA}. Notice a sharp spike in the radial derivative, signalling what we term ``blueshift'' phenomena. {\bf Bottom panel:} the total energy $E$ in the spacetime, defined in Eq.~\eqref{eq:E}. Energy amplification is apparent.}
    \label{fig:scalar}
\end{figure}
Our results are summarized in Fig.~\ref{fig:scalar} and Table~\ref{table:results}. A common feature to all our simulations is clear in Fig.~\ref{fig:scalar}: a sharp spike of radiation, produced close to the inner horizons and released out when the horizons disappear. This is the wave analog of the blueshift seen before in null geodesics, and is responsible for the energy amplification shown in the lower panel of the figure and quantified also in Table~\ref{table:results}. A study of different families of initial data confirms the geodesic estimate~\eqref{amplification_geo}: the amplification ${\cal A}$ scales exponentially with the lifetime of the horizons and with the inner horizon surface gravity. It is also reassuring to notice that the amplification $\mathcal{A}$ of the scalar pulse lies between the largest and smallest values of the corresponding amplification factor of null geodesics, i.e. $\mathcal{A}_g^{\rm min}<\mathcal{A}<\mathcal{A}_g^{\rm max}$ (cf. Table~\ref{table:results}).

The spacetime also possesses one unstable light ring, which corresponds to the maximum real root of the equation $2 F(r) - r F'(r) = 0$ (in the stage where $m$ is constant to a good approximation). Thus, we expect all the physics associated to unstable light rings -- for example, quasinormal ringdown~\cite{Ferrari:1984zz,Cardoso:2008bp,Cardoso:2019rvt} -- to be shared by our dynamical spacetime. The fractional correction to the light-ring frequency is $\delta\Omega/\Omega=2l^2/(27m_0^2)+{\cal O}(l/m_0)^4$, and thus for microscopic $l$ not accessible to current detectors.

\noindent {\bf \em Discussion.} 
%
The novel energy amplification mechanism that we have identified here originates from a blueshift instability which is akin, but nonetheless significantly different, from the Cauchy horizon instability discovered by Penrose~\cite{Penrose:1968ar}. 

Cauchy horizon instabilities have been studied extensively~\cite{Poisson:1989zz,Ori:1991zz,Dafermos:2003wr,Luk:2015qja,Luk:2015pay, Costa:2017tjc,Cardoso:2017soq,Luk:2017jxq,Costa:2019uny,Hollands:2019whz,Zilberman:2022aum}, in particular in the related context  of regular (eternal) black holes~\cite{Brown:2011tv,Carballo-Rubio:2021bpr,Bonanno:2022jjp}.  In all these cases, the inner horizon is an ingoing Cauchy horizon and the corresponding blueshift mechanism leads to a divergent instability and the blow-up of perturbations, which remain confined to the black hole interior. In fact, the global causal structure of these spacetimes is significantly different from the one depicted in Fig.~\ref{fig:penrose} (for comparison, see for instance Fig. 1 in \cite{Bonanno:2022jjp}), and the infinite blueshift instability can already be foreseen by inspection of the corresponding Penrose diagrams: at the level of geometric optics, it stems from the piling up (at the Cauchy horizon) of an infinite number of ingoing null geodesics emitted at regular time intervals, as measured by an exterior observer. 

By contrast, in our framework, the inner horizon {\em is not} a Cauchy horizon; in particular, spacetime is globally hyperbolic and inextendible. Moreover, the inner horizon is outgoing and, in this case, the blueshift instability leads to a finite amplification, in accordance with the fact that now only a finite number of equally timed outgoing geodesics can pile up near the inner horizon. Most remarkably, these phenomena can be observed from infinity, after the horizons evaporate, leading, in essence, to an energy extraction mechanism.

If indeed collapse is halted and horizons disappear, our results seem to imply that there are high-energy phenomena in the cosmos that could be of quantum origin. These effects could give rise to high energy photons, neutrinos or gravitational waves.
Page has shown that primordial black holes of mass $m_0\lesssim 5\times 10^{14}\,{\rm g}$ would have evaporated by now~\cite{Page:1976df}. Take therefore, for illustration purposes, an object which is evaporating today, hence with $v_d-v_c \sim 13\times 10^9\,{\rm yr}$. For Planck size cores, $l\sim 10^{-44}\,{\rm s}$, we find an amplification ${\cal A}\sim e^{10^{61}}$, which means that the outgoing pulse has a significant backreaction in the spacetime, not taken into account in our study. Indeed, a single cosmic microwave background photon would be amplified to a much larger energy than the object itself (and it's challenging if not impossible to concoct an initial fluctuation for which this is not true). Indeed, even for macroscopic $l$ the amplification is tremendous.

Backreaction in the geometry is an interesting problem, but it requires knowledge of an underlying theory leading to spacetime~\eqref{metric}-\eqref{eq:mass_function}. Attempts in this direction can be found in Refs.~\cite{Barcelo:2020mjw,Barcelo:2022gii}.

We did not dwell on more classical phenomena like black hole ringdown, but it is clear that for spacetimes for which the horizons linger longer than a light ring timescale, black hole ringdown should also be observed~\cite{Cardoso:2016rao,Cardoso:2019rvt,Cardoso:2017cqb}.

Finally, our results could have experimental verification, beyond the gravitational realm, in the context of analogue gravity~\cite{1980ApJ...235.1038M,Unruh:1980cg,Visser:1997ux,Barcelo:2005fc}. For example, sound waves in a nontrivial flow propagate as a scalar field on a curved spacetime. For flows with sonic points (where flow velocity equals local sound speed) horizons appear. 
Spherically symmetric, or more general time-dependent flows, therefore develop apparent acoustic horizons and no acoustic event horizons~\cite{Visser:1997ux}. There are indeed acoustic geometries with multiple sonic points~\cite{1997ApL&C..35..389L,Schnerr1994}. It would be interesting to understand the consequences of our results for those acoustic setups, or even if they have a bearing on collapse of air bubbles leading to sonoluminescence~\cite{Brenner:2002zz}.

\noindent {\bf \em Acknowledgments.} 
We are indebted to Ra\'ul Carballo-Rubio for pointing us to important work, of which we were unaware, in a previous version of this manuscript.
V.C.\ is a Villum Investigator and a DNRF Chair, supported by VILLUM Foundation (grant no.\ VIL37766) and the DNRF Chair program (grant no. DNRF162) by the Danish National Research Foundation. V.C.\ acknowledges financial support provided under the European Union's H2020 ERC Advanced Grant ``Black holes: gravitational engines of discovery'' grant agreement
no.\ Gravitas--101052587. Views and opinions expressed are however those of the author only and do not necessarily reflect those of the European Union or the European Research Council. Neither the European Union nor the granting authority can be held responsible for them.
This project has received funding from the European Union's Horizon 2020 research and innovation programme under the Marie Sklodowska-Curie grant agreement No 101007855.
We acknowledge financial support provided by FCT/Portugal through grants 
2022.01324.PTDC, PTDC/FIS-AST/7002/2020, UIDB/00099/2020 and UIDB/04459/2020.

Z.Z.\ acknowledges financial support from China Scholarship Council (No.~202106040037).

J.L.C and J.N. were partially supported by FCT/Portugal through CAMGSD, IST-ID ,
projects UIDB/04459/2020 and UIDP/04459/2020, and by FCT/Portugal and CERN through project CERN/FIS-PAR/0023/2019 

\bibliography{ref}

\end{document}